# Energy recovery from *Ginkgo biloba* urban pruning wastes: pyrolysis optimization and fuel property enhancement for high-grade charcoal productions

**Padam Prasad Paudel,**  Department of Biosystems Engineering, Kangwon National University, Chuncheon-si, Republic of Korea; Department of Soil Science and Agri-Engineering, Faculty of Agriculture, Agriculture and Forestry University, Chitwan, Nepal
**Sunyong Park, Kwang Cheol Oh, Seok Jun Kim,** Agriculture and Life Science Research Institute, Kangwon National University, Chuncheon-si, Republic of Korea
**Seon Yeop Kim, Kyeong Sik Kang,** Department of Interdisciplinary Program in Smart Agriculture, Kangwon National University, Chuncheon-si, Republic of Korea
**Dae Hyun Kim,**  Department of Biosystems Engineering, Kangwon National University, Chuncheon-si, Republic of Korea; Agriculture and Life Science Research Institute, Kangwon National University, Chuncheon-si, Republic of Korea; Department of Interdisciplinary Program in Smart Agriculture, Kangwon National University, Chuncheon-si, Republic of Korea

Abstract: *Ginkgo biloba* trees are widely planted in urban areas of developed countries for their resilience, longevity and aesthetic appeal. Annual pruning to control tree size, shape and interference with traffic and pedestrians generates large volumes of unutilized Ginkgo biomass. This study aimed to valorize these pruning residues into charcoal by optimizing pyrolysis conditions and evaluating its fuel properties. The pyrolysis experiment was conducted at 400–600°C, after oven drying pretreatment. The mass yield of charcoal was found to vary from 27.33 to 32.05% and the approximate volume shrinkage was found to be 41.19–49.97%. The fuel properties of the charcoals were evaluated using the moisture absorption test, proximate and ultimate analysis, thermogravimetry, calorimetry and inductively coupled plasma optical emission spectrometry. The calorific value improved from 20.76 to 34.26 MJ kg$^{-1}$ with energy yield up to 46.75%. Charcoal exhibited superior thermal stability and better combustion performance. The results revealed satisfactory properties compared with other biomass, coal and biochar standards. The product complied with first-grade standards at 550 and 600°C and second-grade wood charcoal standards at other temperatures. However, higher concentrations of some heavy metals like Zn indicate the need

Correspondence to: Dae Hyun Kim, Department of Biosystems Engineering, Kangwon National University, Hyoja 2 Dong 192-1, Chuncheon-si, Republic of Korea. E-mail: daekim@kangwon.ac.kr

for pretreatment and further research on co-pyrolysis for resource optimization. This study highlights the dual benefits of waste management and renewable energy, providing insights for urban planning and policymaking. © 2025 The Author(s). *Biofuels, Bioproducts and Biorefining* published by Society of Industrial Chemistry and John Wiley & Sons Ltd.

## Introduction

*Ginkgo biloba*, a living fossil with a history dating back 280 million years, is widely cultivated in urban areas across developed countries for its ornamental value, adaptability to urban conditions, and resistance to pests, diseases and pollution.[1,2] In cities like Seoul and Tokyo, *G. biloba* constitutes a significant proportion of urban forests, contributing to shade, aesthetics and ecological balance.[3,4] Introduced to Japan 900 years ago and to Europe and North America in the early eighteenth century, its genetic origins trace back to Korean accessions.[1,5] The tree's resilience to urban stressors, such as salt stress, and its ability to intercept rainfall and enhance the urban water cycle, highlight its ecological importance.[6–8] Its leaves, rich in polysaccharides and bioactive compounds with antioxidant, anti-inflammatory, and neuroprotective properties, have been widely utilized in herbal medicine, further increasing its value.[9,10] While its urban benefits are significant, considerations such as allergic reactions to pollen and waste management challenges must be addressed in sustainable urban planning.[3,11] However, pruning waste from Ginkgo trees, particularly from annual or biannual maintenance to manage size, aesthetics and public safety, remains largely underutilized. In China alone, 40 000 tons of Ginkgo waste are discarded annually, contributing to environmental pollution and carbon emissions.[11] This underscores the need for technologies that recycle pruning waste, transforming it into a valuable resource and reducing improper disposal practices.

In South Korea, Ginkgo trees represent a significant portion of urban greenery. Out of 294 668 street trees in Seoul, 102 070 are Ginkgos, accounting for 35.64% of the total, thereby resulting in approximately 17.17 Ginkgo trees per square kilometer across the city's 605 km$^2$.[12] Assuming an adult deciduous plant like Ginkgo in temperate region can yield an average of 10 kg of pruning biomass annually,[13–15] the total Ginkgo pruning biomass in Seoul is estimated at 1 020 700 kg per year. Extending this figure to the entire urban tree population across South Korea's total urban area of 17 590 km$^2$,[16] the total Ginkgo pruning biomass generated annually would reach 29 676.22 metric tons. With an energy content of 20.76 MJ per kg, this corresponds to a potential energy generation of 616.08 terajoules per year. While this energy potential is minimal in comparison with South Korea's annual energy production of 617 TWh,[17] utilizing Ginkgo pruning biomass could offer a valuable avenue for waste utilization in urban environments.

Extensive literature research reveals no studies specifically addressing the utilization of *G. biloba* pruning (GBP) waste for energy or fuel production. However, related studies have explored various applications of *G. biloba* biomass, including leaves and seeds. Yu *et al.*[18] investigated the pyrolysis of *G. biloba* leaves, converting them into aromatic and high-quality hydrocarbons for bio-oil and gas production. Similarly, Zhang *et al.*[19] conducted co-pyrolysis of *G. biloba* leaves with rubber, enhancing the specific surface area of the resulting biochar to 337.50 m$^2$/g at 900°C. Wang *et al.*[20] identified valuable bioactive components in the seed exocarp of *G. biloba*, which can be extracted through bioprocessing. Additionally, Wang *et al.*[21] developed an enzyme system achieving an 87.2% conversion rate of *G. biloba* leaves into reducing sugars, highlighting its potential for bio-based industries. Zhang *et al.*[22] further examined the thermal stability and kinetics of *G. biloba* leaves, including those subjected to various extraction processes. While no direct studies focus on the energy or fuel conversion of GBP, these findings on biomass processing and waste utilization suggest promising avenues for exploration.

Unlike those herbaceous urban biomass streams (e.g. leaf litter or grass clippings) wastes, Ginkgo pruning residues (branches and stems) are woody in nature and fundamentally different, with a higher lignocellulosic content, particularly rich in lignin and cellulose (cellulose ~42–45%, lignin ~27–31% and hemicellulose ~10–22%).[23,24] Compared with other hardwood residues, this composition makes them more suitable for producing high-quality charcoal, as the lignocellulosic material can yield a product with higher carbon content but requiring more robust pyrolysis conditions.[25] Urban pruning waste also poses challenges, such as potential contamination with heavy metals, dust and vehicle exhaust particulates. Thus, pruning residues present both challenges and opportunities for valorization.

The thermochemical conversion process is a promising approach for managing biomass waste and producing value-added products.[26] Pyrolysis, a key thermochemical route, transforms biomass into syngas, bio-oil and biochar, each with significant potential for carbon sequestration and as renewable energy alternatives. This process contributes to climate change mitigation, waste management, the addressing of energy crises and the promotion of sustainable practices.[25] Among the products, biochar (or charcoal) stands out for its versatile applications, making it a valuable resource for addressing environmental challenges across energy, industrial, and agricultural sectors. In agriculture, biochar serves as a soil amendment, enhancing water retention, nutrient absorption and carbon sequestration. In the energy sector, it actss as a renewable fuel, substituting for coal in power plants and industrial processes, with studies highlighting its potential for co-firing with coal and as a fuel for heat generation.[27]

The primary objective of this paper was to convert GBP into charcoal, for potential use as a solid fuel. Fuel property evaluations and compliance with the standards and regulation were subsequent objectives, whereas optimizing the pyrolysis process for better quality and higher yield was the secondary objective. The study stands out for its focus on a novel feedstock–*G. biloba* pruning–which has not been extensively explored for energy applications. Key aspects of this research include the proximate analysis, chemical composition, thermal stability and energy release potential of Ginkgo branches; process optimization; compliance with fuel standards; and an evaluation of environmental benefits and sustainability. Additionally, this study highlights the potential of utilizing urban tree waste as an energy resource, contributing to waste-to-energy solutions.

## Materials and methods

### Feedstock preparation

Pruned Ginkgo tree branches were collected from the premises of Kangwon National University, South Korea, and cut transversely into cylindrical disc-shaped chips. The bark was removed to ensure sample homogeneity for consistent analysis. The chips were then dried in an oven at 105°C for 72 h, resulting in a moisture content of 50.67 ± 6.42% on a mass basis. After drying, the chips had an average thickness of 26.3 ± 1.95 mm, with diameters ranging from 49.42 to 72.39 mm. A total of 30 chips were prepared, each uniquely labeled.

### Pyrolysis experiments

The pyrolysis was conducted in an electric furnace (model N7/H/B410, Nabertherm GmbH, Germany) that was preheated to target temperatures of 400, 450, 500, 550 and 600°C. Once the furnace had stabilized at the desired temperature, the samples were quickly placed inside to achieve rapid heating and maintained at the target temperature for 1 h. The temperature range 400–600°C was selected for its proven effectiveness in maximizing charcoal yield, fixed carbon content and energy density for high-quality fuel production, as temperatures below 400°C result in incomplete carbonization and temperatures above 600°C significantly decrease charcoal yield with minimal improvement in carbon quality.[28,29] Previous work by Syred *et al.*[28] describes woody biomass pyrolysis as a three-stage process of moisture removal, decomposition (110–270°C), and carbonization (400–600°C), indicating that the bulk of fixed-carbon formation occurs within the chosen window, a finding further confirmed by thermogravimetric analysis. Similarly, Adeniyi *et al.*[29] report peak anhydrous weight yields at 400°C and a marked decrease above 600°C owing to secondary reactions. For each batch, six Ginkgo chips were placed in three stainless steel cans (diameter 85 mm, height 80 mm), with two chips of varying sizes in each can to minimize mass variability. These cans were equipped with lids that allowed the emission of volatile matter while preventing the intake of outside air, creating a self-generated atmosphere that ensured an oxygen-free environment. This approach was selected to prioritize solid yield, simplify the experimental setup and reduce costs, since using an inert gas such as $N_2$ as a sweep has been shown to enhance volatile (gas and oil) production at the expense of solid products.[25] The resulting charcoal products from each temperature condition were labeled as G400, G450, G500, G550 and G600, respectively, and the dried unprocessed feedstock was lebelled as RAW.

### Laboratory analysis and measurements

#### Mass and dimensions measurements

The mass of the Ginkgo chips was measured using a precision weighing machine with an accuracy of 0.01 g. The dimensions, including height and diameter, were measured using a precision vernier caliper with a least count of 0.01 mm. The average diameter ($D_{avg}$) of each chip was determined by averaging the major (maximum) and minor (minimum) diameters at the middle girth of the chip.

The mass and geometric dimensions of the chips were recorded before and after pyrolysis. The approximate volume ($V$) of each chip was calculated using Equation (1):

$$V = \frac{\pi D_{avg}^2}{4} \times H \quad (1)$$

where $H$ is the height of the chip.

The density ($\rho$) of the chips was calculated as the ratio of mass ($M$) to the approximate volume ($V$) using Equation (2):

$$\rho = \frac{M}{V} \quad (2)$$

The mass yield (MY) and mass reduction (MR) after pyrolysis were calculated using Equations (3) and (4), respectively:[30]

$$\text{MY} = \frac{M_2}{M_1} \times 100 \quad (3)$$

$$\text{MR} = \frac{M_1 - M_2}{M_1} \times 100 = 100 - \text{MY} \quad (4)$$

where $M_1$ is the initial mass and $M_2$ is the final mass after pyrolysis.

The linear shrinkage for any one-dimensional measurement $L$ (either average diameter $D$ or height $H$) was calculated using Equation (5):

$$\text{Linear shrinkage} = \frac{L_1 - L_2}{L_1} \times 100 \quad (5)$$

where $L_1$ and $L_2$ are the initial and post-pyrolysis value of the chosen dimension. Equation (5) was applied separately to $D$ and $H$ to quantify both diameter and height reduction, allowing the observed anisotropic shrinkage to be captured.

The volume reduction was calculated similarly using Equation (6):

$$\text{Volume reduction} = \frac{V_1 - V_2}{V_1} \times 100 \quad (6)$$

where $V_1$ and $V_2$ are the initial and final volumes, respectively.

The density reduction was determined using Equation (7):

$$\text{Density reduction} = \frac{\rho_1 - \rho_2}{\rho_1} \times 100 \quad (7)$$

where $\rho_1$ and $\rho_2$ are the initial and final densities, respectively.

### Moisture absorption test and contact angle measurement

The moisture absorption test was conducted by exposing the oven-dried pyrolyzed product to a typical environment in a growth room having temperature 27°C and relative humidity between 70 and 75%. The duration of exposure was 72 h. The amount of moisture absorbed (MA, %) was calculated as using Equation (8):

$$\text{MA}(\%) = \frac{M_f - M_i}{M_i} \times 100 \quad (8)$$

where $M_i$ and $M_f$ are the masses of charcoal before and after moisture absorption test (g), respectively.

Additionally, the hydrophobicity of Ginkgo charcoals was evaluated through contact angle measurements using a Surface Electro Optics contact angle goniometer, Phoenix-MT(M) model.

### Calorimetry

Following oven drying, the higher heating values (HHVs) of each sample were measured in triplicate using a CAL-3K calorimeter (DDS Calorimeters, South Africa). The energy yield (EY, %) was subsequently calculated based on the mass yield and HHV of both charcoal and raw feedstock, as shown in Equation (9):

$$\text{EY}(\%) = \frac{\text{HHV}_{\text{Charcoal}}}{\text{HHV}_{\text{Raw}}} \times \text{MY} \quad (9)$$

### Elemental analysis

The percentages of carbon (C), hydrogen (H) and nitrogen (N) in biomass and charcoal were analyzed using an elemental analyzer (FlashEA 1112, Thermo Fisher Scientific, Waltham, MA, USA). The oxygen (O) content was calculated by difference using Equation (10):

$$\text{O}(\%) = 100 - [\text{C}(\%) + \text{H}(\%) + \text{N}(\%)] \quad (10)$$

A Van Krevelen diagram was developed based on the H/C and O/C atomic ratios to assess the chemical characteristics and degree of aromaticity of the samples.

### Proximate analysis

The moisture content (MC), volatile matter (VM) and ash content (AC) were determined using a proximate analyzer (Precisa PrepASH 229, Switzerland) following ASTM D7582-12 standards. The fixed carbon (FC) was calculated by difference using Equation (11):

$$\text{FC}(\%) = 100 - [\text{VM}(\%) + \text{AC}(\%) + \text{MC}(\%)] \quad (11)$$

Based on proximate analysis, the fuel ratio (FR), calculated as the ratio of FC to VM percentages, is an important metric in power plants, typically ranging from 0.5 to 3.0. It plays a key role in defining combustion properties. However, if the FR exceeds 2.0, ignition and flame stability issues may arise, potentially complicating the combustion process.[31]

As the combustion indices, combustibility index (CI) and volatile ignitability (VI), are evaluated to access the pyrolysis efficiency and fuel quality. The CI is crucial for thermal power plants and coal blending:

$$CI = \frac{HHV}{FR} \times (115 - Ash) \times \frac{1}{115} \qquad (12)$$

$$VI = \frac{HHV_{db} - 0.338 FC_{db}}{VM_{db} + M} \times 100 \qquad (13)$$

The VI quantifies the potential energy from volatiles, assuming the fixed carbon is pure carbon. For optimal VI, the volatile matter should have a specific HHV of at least 14 MJ kg$^{-1}$. The VI is expressed by Equation (12), where HHV is the higher calorific value (MJ kg$^{-1}$), and MC represents moisture content (%), with all values calculated on a dry basis.[31,32]

### Thermogravimetric analysis

Thermogravimetric analysis (TGA) of the raw *G. biloba* branch samples and their charcoals was performed using a simultaneous thermal analyzer (model SDT650, TA Instruments, Delaware, USA). The analysis was conducted under both the nitrogen atmosphere and air conditions, with a flow rate of 100 mL/min. The temperature program included an initial isothermal hold at 35°C for 5 min, followed by a linear heating ramp at 10°C/min up to 900°C from room temperature. The weight loss of the sample was recorded continuously as a function of temperature, allowing for the assessment of decomposition stages and thermal stability.

### Heavy metal and mercury analysis

To access the environmental hazard potential of Ginkgo charcoal, heavy metal and mercury concentrations were analyzed to ensure compliance with safety standards for fuel use. Mercury content was determined using a mercury analyzer (Teledyne Leeman Labs Hydra II C). For the analysis of other heavy metals, including As, Cd, Cr, Cu, Ni, Pb and Zn, samples were pretreated in a microwave digestion system, followed by quantification using an inductively coupled plasma optical emission spectrometer (Agilent 5900, Agilent Technologies).

## Results and discussions

### Mass and dimensions reductions

As the pyrolysis temperature increases from 400 to 600°C, the biomass exhibits significant reductions in mass, volume, height, diameter and density as illustrated in Fig. 1. The mass reduction progressively rises from 67.96 to 73.60%, indicating enhanced decomposition and volatilization at higher temperatures. Volume reduction also increases from 41.19 to 49.97%, demonstrating substantial shrinkage owing to the breakdown of biomass structure. However, the density reduction demonstrates relative stability across these temperatures, indicating a proportional relationship between mass loss and volume reduction. Height reduction increases notably from 16.07 to 26.07%, while the diameter reduction remains relatively stable, ranging from 16.30 to 17.74%, suggesting an anisotropic shrinkage pattern with greater vertical compaction than lateral. This can be attributed to the longitudinal alignment of cell fibers and differential moisture loss along the grain direction. These trends indicate that higher pyrolysis temperatures significantly affect the physical dimensions and mass of the biomass, influencing the resulting charcoal's structural and physical properties. This significant reduction in mass and volume indicates a highly energy densified product and facilitates convenient transportation and space saving in storage.[33]

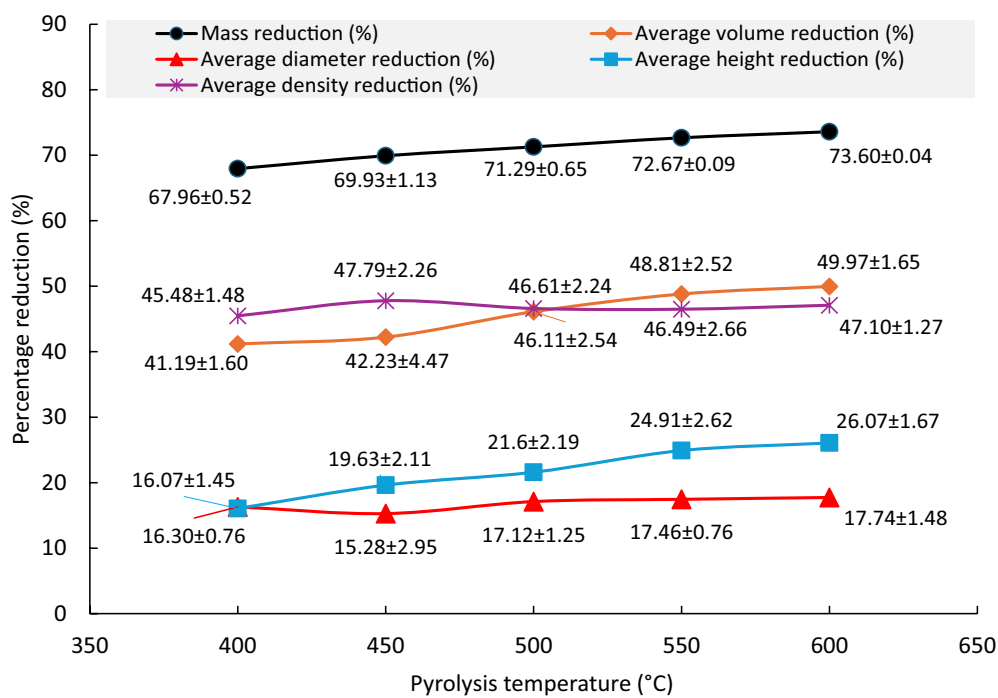

Figure 1. Reduction of size and mass along with pyrolysis temperature.

## Moisture absorptions and hydrophobicity

Pyrolysis temperature significantly influenced the moisture absorption properties of charcoal, as interpreted in Fig. 2. The amount of moisture absorption was found to be negatively correlated with pyrolysis temperature with an $R^2$ of 0.971, where the highest absorption was recorded at 9.04% for G400 Ginkgo charcoal and the lowest at 5.90% for G600 charcoal. This result is within the normal range as charcoal can absorb between 3 and 12% moisture at higher relative humidity levels.[34] The amount of moisture absorption depends on the pyrolysis temperature, environmental exposure, storage conditions, exposed time and feedstock types.[35,36] Generally, lower temperatures yields have hydrophilic functional groups (e.g. hydroxyl and carboxyl groups) and potential use for soil amendment,[37] while higher temperatures (≥400°C) have functional groups that favor hydrophobicity,[38] enhancing storage stability, and have better application prospects for fuel purposes.

Contact angle measurements further demonstrated the impact of pyrolysis temperature on the hydrophobicity of charcoal surfaces. The contact angle increased from 95.6° in raw biomass to 108.31° at 400°C, reaching a peak value of 114.37° at 450°C, indicating increased hydrophobicity owing to the decomposition of hydrophilic functional groups and increased aromatic carbon content.[38] Beyond 500°C, contact angle values plateaued at 110.63° (500°C), 110.29° (550°C) and 110.96° (600°C), suggesting that while chemical hydrophobicity continued to improve, the influence of porosity and surface roughness became more significant at higher temperatures, which promotes water penetration.[39] This was further confirmed by the anisotropic water absorption behavior, where the axial surfaces (plain, top and bottom) absorbed water more quickly than the curved transverse surfaces. This observation, consistent with the water absorption results, highlights a trade-off between chemical hydrophobicity and physical porosity. Charcoal produced at higher pyrolysis temperatures showed increased resistance to environmental moisture absorption but readily absorbs and releases water when in direct contact with liquid, emphasizing the dual role of high-temperature charcoal as both hydrophobic and porous.[40,41]

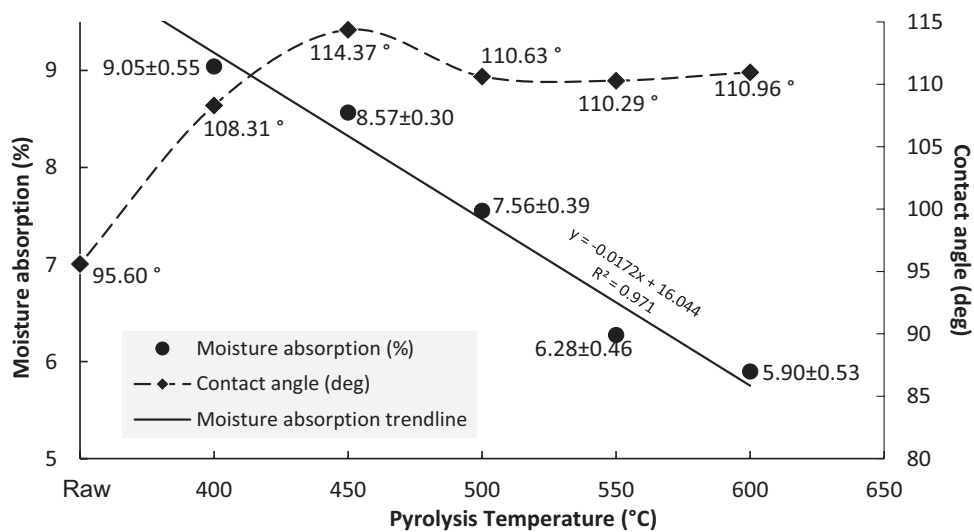

Figure 2. Moisture absorption pattern and hydrophobicity with process temperature.

## Elemental analysis and proximate analysis

### Elemental analysis

The results of the elemental analysis of raw Ginkgo and its charcoal products are presented in Fig. 3. The carbon (C) content increased significantly from 48.33% (raw) to 85.61% (G600) with a clear increasing trend as rising the pyrolysis temperature, whereas there was a decrease in oxygen (O) and hydrogen (H) content. This increasing carbonization is due to progressive thermal decomposition of volatile compounds and the concentration of fixed carbon as the temperature rises.[42] The decline in oxygen content from 45.87% (raw) to 11.67% (G600) is due to the dehydration and decarboxylation with releasing oxygen in the forms of CO, $CO_2$, and water vapor.[43] Dehydration and volatilization of hydrogen containing volatile compounds such as hydrocarbons are attributed to declining hydrogen content,[44] from 5.67% (raw) to 2.35% (G600). In addition, the relative stability in nitrogen (N) content across all charcoals found with minimal variation (approximately 0.3–0.37%) is probably due to the limited volatilization of nitrogen compounds compared with other elements like hydrogen and oxygen.[45] These decreased hydrogen and oxygen contents and increased carbon content enhance the features for solid fuel applications and greater thermal stability.

The Van Krevelen diagram (Fig. 4) illustrates the relationship between the atomic ratios of C, H and O. The raw feedstock exhibited the highest H/C and O/C ratios, while when the processing temperature increased, the charcoal samples showed systematic decreases in both ratios with a strong linear correlation ($R^2 = 0.9988$) between the two ratios. This highlighted the consistency of compositional changes during thermal treatment. The clustering of points near the origin indicated the greater thermal stability, higher heating value and increased aromaticity and carbon content of Ginkgo charcoals.

### Proximate analysis

The proximate composition along with fuel ratio of raw Ginkgo and its charcoals are presented in Fig. 5. As the pyrolysis temperature increased, both MC and VM decreased significantly, while FC and FR increased, indicating improved carbonization and energy density. Specifically, VM decreased from 74.5% in RAW to 12.56% in G600, while FC rose from 21% to 79.95%, resulting in a fuel ratio increase from 0.28 to 6.36. These changes align with the elemental composition discussed earlier, highlighting the incremental removal of volatile compounds and the concentration of fixed carbon at higher pyrolysis temperatures. The slight increase in ash content is due to the thermal degradation of organic matter and concentration of inorganic minerals.[46] The improved FR for higher temperatures indicates a shift toward more fixed carbon relative to volatile matter. High-temperature charcoals, such as G550 and G600, are more suitable for energy applications owing to their high FC and fuel ratio, which indicate better combustion properties.

## Thermogravimetric analysis

Figure 6 depicts the TGA results of raw samples and charcoals of Ginkgo under nitrogen ($N_2$) and air atmospheres,

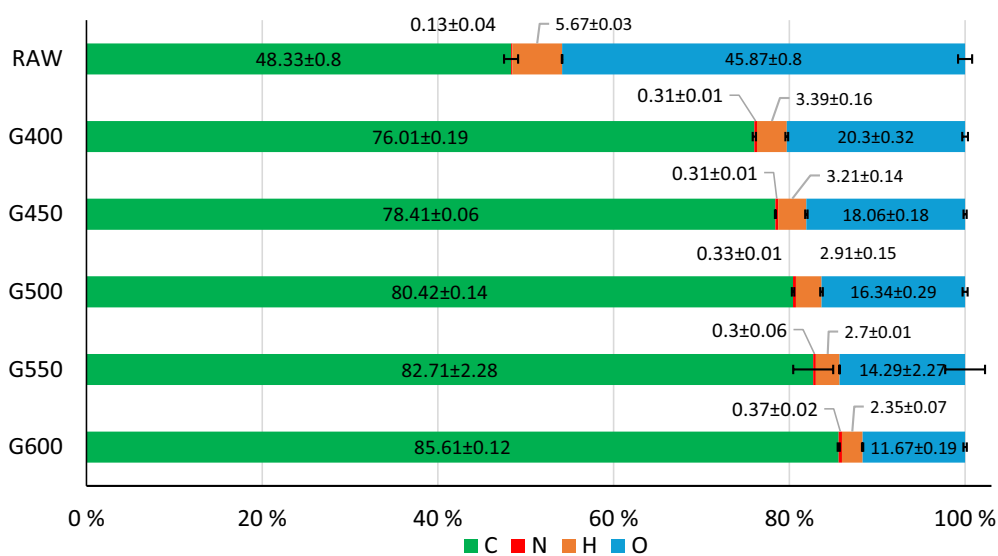

Figure 3. Composition of C, N, H, O (elemental analysis).

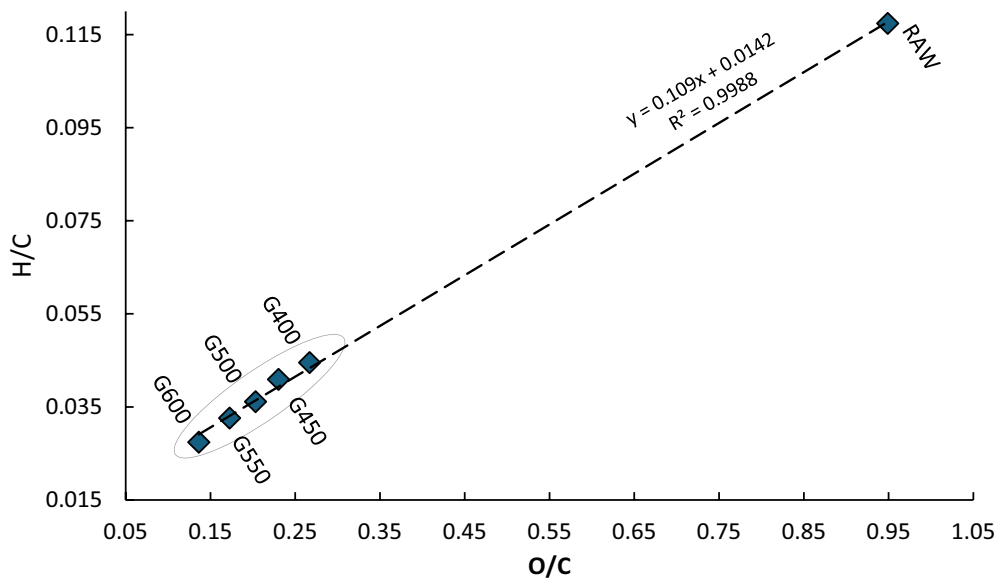

Figure 4. Van Krevelen diagram.

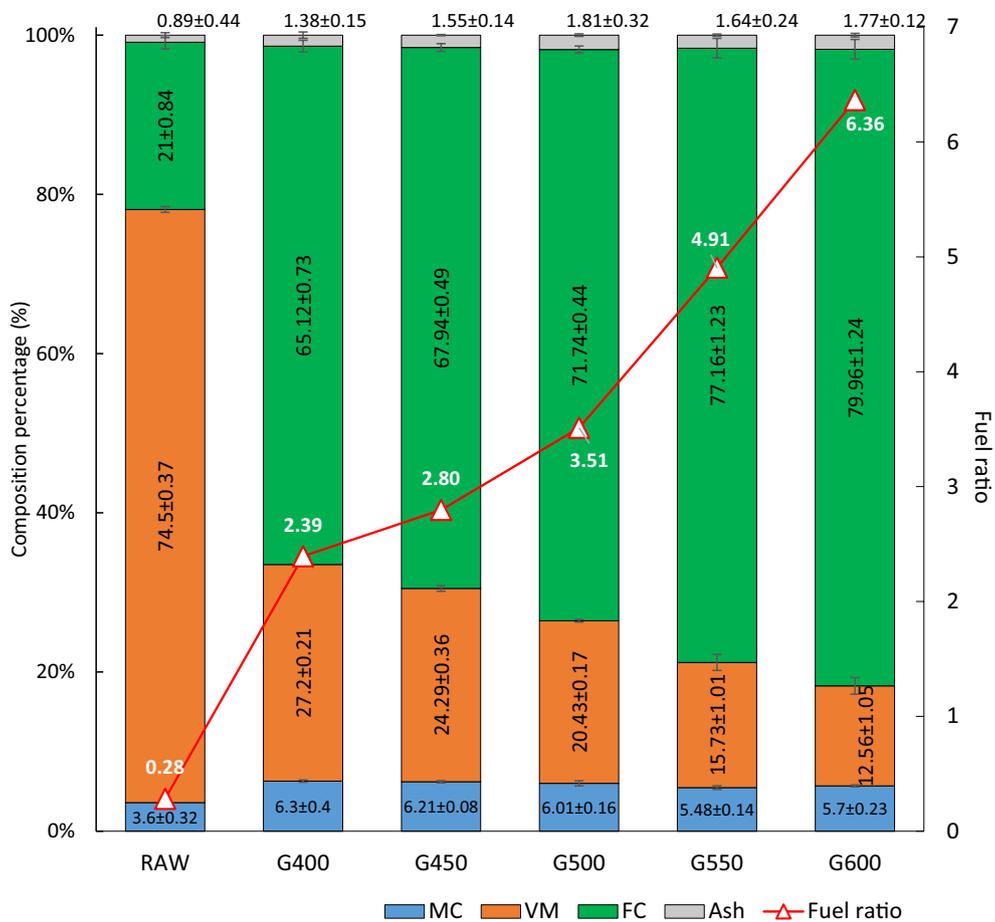

Figure 5. Composition of moisture, volatile matter, ash and fixed carbon.

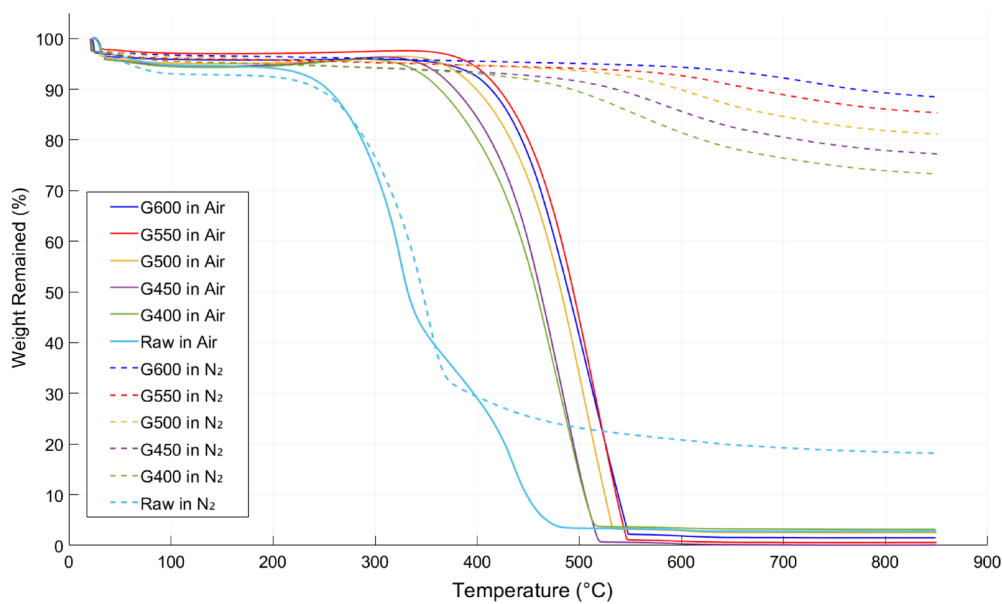

Figure 6. Thermogravimetric analysis results under the $N_2$ medium and air medium.

illustrating the weight loss profiles and thermal degradation behavior as a function of temperature. The TGA of the raw sample revealed four major phases: the dehydration phase (below 200°C), where moisture evaporates; the primary decomposition phase (200–350°C), characterized by the release of volatile matter; the secondary decomposition phase (350–500°C), involving the combustion of fixed carbon; and the residual mass phase (above 500°C), where ash forms after oxidation particularly under the air condition. These observations align with the typical wood decomposition behaviors, where moisture evaporation, hemicellulose and cellulose decomposition, and lignin decompositions are the main phases, followed by stable ash formations.[47,48]

For the charcoal sample, three main phases were observed under the air condition. The first phase involved dehydration and relative mass stability with about 5% mass loss below 350°C. The second phase, between 350 and 550°C, shows a combustion phase with more than 90% mass loss under the air condition. The final phase involves less than 3% residual ash accumulation under the air condition. The absence of secondary decomposition is due to the limited amount of volatile matter for combustion. In inert ($N_2$) conditions, the mass remained stable below 400°C, with slight mass loss occurring between 400 and 800°C. However, a higher residual mass of 73–89% was observed owing to the retention of fixed carbon.

As the pyrolysis temperature increased, the mass loss decreased, leading to an upward shift in the shape of the graph for $N_2$, while a mixed pattern was observed under the air condition for Ginkgo charcoal. The TGA reveals that increasing the carbonization temperature produces progressively more thermally stable Ginkgo chars. In nitrogen, G600 exhibited a smaller mass loss and a delayed onset of major decomposition compared with G400, reflecting its lower volatile content and higher fixed-carbon fraction. The G600 charcoal held the greatest thermal stability in the $N_2$ condition, whereas G550 exhibited better thermal stability and lower residual mass under the air condition as G550 charcoal began to oxidize at a higher temperature and degraded more sharply than the other charcoals. The superior combustion performance of G550 compared with G600 can be linked to the requirement for a significant amount VM that promotes pre-heating followed by fast fixed-carbon oxidation.[30,49,50] The ignition start temperature was around 350°C for Ginkgo charcoals, compared with around 250°C for raw wood. The TGA profile of charcoals showed a well-carbonized form with low volatile matter and high carbon stability, characterizing its suitability for high thermal resilience applications.

## Mass yield, calorific value, energy yield and indices

The effects of thermal treatment and fuel properties are presented in Table 1 in terms of MY, HHV, EY, CI and VI. The HHV significantly improved from 20.76 MJ kg$^{-1}$ in the raw sample to a maximum of 34.26 MJ kg$^{-1}$ in the G600 sample, indicating a 65% improvement in energy density. Concurrently, a notable decrease in MY was observed to 26.41%, and in EY to 43.58% for G600. Both MY and EY

**Table 1. Mass yield (MY), high heat value (HHV), energy yield (EY) and indices.**

| Sample | HHV (MJ kg$^{-1}$) | MY (%) | EY (%) | FR | CI | VI |
|---|---|---|---|---|---|---|
| RAW | 20.76 ± 0.50 | 100.00 | 100.00 | 0.28 | 80.02 | 17.49 |
| G400 | 30.28 ± 0.27 | 32.05 ± 0.51 | 46.75 | 2.39 | 13.69 | 24.68 |
| G450 | 30.97 ± 0.26 | 30.08 ± 1.12 | 44.87 | 2.80 | 11.97 | 26.25 |
| G500 | 32.48 ± 0.28 | 28.72 ± 0.65 | 44.94 | 3.51 | 9.97 | 31.13 |
| G550 | 33.17 ± 0.43 | 27.33 ± 0.09 | 43.68 | 4.91 | 7.30 | 33.44 |
| G600 | 34.26 ± 0.28 | 26.41 ± 0.03 | 43.58 | 6.36 | 5.80 | 39.61 |

CI, Combustibility index; FR, fuel ratio; VI, volatile ignitability.

decreased with increasing temperature treatment, while the HHV increased.

The value of CI for raw samples was 80.02 MJ kg$^{-1}$, then after pyrolysis it abruptly decreased to 13.69 MJ kg$^{-1}$ for G400 and then gradually decreased to 5.80 MJ kg$^{-1}$ for G600. Lower values of CI may lead to slower combustion performance and consumption of less mass during the combustion process owing to the lower volatile content.[51,52] Meanwhile there was a gradual increase in VI from 17.49 MJ kg$^{-1}$ for raw to 39.61 MJ kg$^{-1}$ for G600 that indicates that the material become less volatile as it undergoes higher temperature treatments. All of these samples exceeded the minimum threshold VI of 14 MJ kg$^{-1}$ for boiler operation and power plants.[53]

## Heavy metal and trace element concentrations

Table 2 presents the inductively coupled plasma optical emission spectrometry and mercury analysis results, highlighting the concentrations of heavy metals and trace elements in Ginkgo charcoal. Cadmium (Cd) and lead (Pb) were consistently below detection levels across all samples. However, arsenic (As), chromium (Cr), copper (Cu), nickel (Ni) and mercury (Hg) exceeded regulatory limits in certain samples. Zinc (Zn), with concentrations ranging from 423.08 to 984.02 mg L$^{-1}$ for all samples, exceeded ISO regulatory limits for thermally treated wood pallets.[54] Among the samples, only G550 charcoal satisfied the Korean regulatory standards for heavy metal and trace element concentrations.[55] The higher concentrations of heavy metals may be attributed to urban environmental factors, such as atmospheric pollution, vehicular exhaust, industrial emissions, closeness to laboratory waste disposal sites and contaminated soil and groundwater, all of which probably contribute to metal uptake. Research demonstrates that street dust in commercial zones can contain up to 1.6 g kg$^{-1}$ of zinc.[56] *Ginkgo biloba*'s strong tolerance to heavy metals and urban pollution may explain its ability to accumulate over 1 g kg$^{-1}$.[57]

During charcoal combustion, most heavy metals either remain concentrated in the ash or volatilize into the flue gas. Elevated Zn and Hg levels can increase metal-laden particulate emissions and cause partial mercury volatilization, potentially exceeding environmental emission guidelines. To comply with regulations, such as the US Environmental Protection Agency limit of <1.2 lb Hg per TBtu and ≤0.010 lb PM$_{10}$ per MMBtu for non-mercury metal hazards in coal-fired power plants, these emissions must be carefully quantified and controlled.[58] After combustion, ash concentrates residual heavy metals and must be evaluated using procedures like the US Resource Conservation and Recovery Act's Toxicity Characteristic Leaching Procedure limit of 5 mg L$^{-1}$ for most heavy metals.[59]

To ensure safe use, Ginkgo charcoals may require pretreatment methods such as mixing with additives, or acid or water washing.[60,61] For instance, using 0.1–1 mol L$^{-1}$ HCl under continuous stirring for 3 h effectively reduces the heavy metal content, specifically iron, by about 77% in charcoal,[62] with up to 99% zinc removal expected.[63] Similarly, 2.47 mol L$^{-1}$ H$_3$PO$_3$ can remove 95% of Zn in 10 h.[59] Blending 25–75% clean biomass such as agricultural residues significantly dilutes the heavy metal concentration by up to 93%.[64] Alternatively, charcoal could be utilized in applications like power generation, where regulatory limits are less stringent, or in isolated areas to minimize human exposure. Robust flue-gas cleaning systems, enhanced exhaust configurations and controlled combustion methods can also reduce heavy-metal emissions in the flue gas.[65] Since most heavy metals remain in the ash during combustion, proper ash disposal, either as hazardous waste under local regulations or as a feedstock for non-food applications like construction materials, is essential to prevent environmental and health risks.[66,67]

## Fuel property optimization and comparison with the fuel standards

Optimization depends on the choice of the parameter and the application purpose. If MY and EY are prioritized, G400

| Table 2. Concentration of heavy metals and trace elements in Ginkgo charcoal. | | | | | | | | | | |
|---|---|---|---|---|---|---|---|---|---|---|
| Metal | G600 | G550 | G500 | G450 | G400 | RAW | Korean standard for wood charcoal (mg kg$^{-1}$) | ISO/TS 17225-8: 2016(E) (mg kg$^{-1}$) | | |
| | | | | | | | | TW1 | TW2 | TW3 |
| As, mg L$^{-1}$ | 2.40 | BDL | 2.19 | BDL | BDL | BDL | ≤1 | ≤1 | ≤2 | ≤2 |
| Cd, mg L$^{-1}$ | BDL | BDL | BDL | BDL | BDL | BDL | ≤1.5 | ≤0.5 | ≤1 | ≤2 |
| Cr, mg L$^{-1}$ | 7.21 | BDL | BDL | BDL | BDL | 17.88 | – | ≤10 | ≤15 | ≤15 |
| Cu, mg L$^{-1}$ | 7.21 | 5.00 | BDL | 4.74 | 4.21 | 4.47 | – | ≤10 | ≤20 | ≤20 |
| Ni, mg L$^{-1}$ | BDL | 5.00 | BDL | BDL | BDL | 6.71 | – | ≤10 | ≤10 | ≤10 |
| Pb, mg L$^{-1}$ | BDL | BDL | BDL | BDL | BDL | BDL | ≤30.0 | ≤10 | ≤10 | ≤10 |
| Zn, mg L$^{-1}$ | 423.08 | 482.50 | 342.26 | 640.11 | 984.02 | 404.56 | – | ≤100 | ≤100 | ≤100 |
| Hg, mg kg$^{-1}$ | 0.19 | BDL | 2.89 | 0.36 | 0.69 | 3.96 | ≤0.15 | ≤0.1 | ≤0.1 | ≤0.1 |

*Note*: BDL, below detection level; As, arsenic; Cd, cadmium; Cr, chromium; Ni, nickel; Pb, lead; Zn, zinc; Hg, mercury; TW1, TW2, TW3, classes of charcoal. The measurement mg kg$^{-1}$ is on a dry basis.

or G450 are optimal. G400 had the highest MY (32.05%) and EY (46.75%) among all charcoals with significantly improved fuel quality compared with RAW. However, if the goal is to produce high-quality charcoal for fuel, G600 performed best, offering the highest energy density (34.26 MJ kg$^{-1}$), least moisture absorption, highest carbon content, lowest volatile matter, superior thermal stability, highest FR (6.36) and VI, and the lowest CI (5.8). However, it had significantly lower yields (MY, 26.41%; EY, 43.58%). Notably, the energy yield difference between G400 and G600 was minimal. Charcoals produced at higher temperatures (e.g. G600 and G550) showed better thermal stability and combustion performance. Based on TGA analysis, G550 in air displayed the steepest weight loss curve during active combustion and the least residual weight, demonstrating superior combustion performance. This makes G550 ideal for applications requiring high energy output and minimal residue, balancing fuel efficiency with yield. If the priority includes minimizing heavy metals and trace elements, G550 is the safest among all the charcoal samples, containing the fewest harmful elements and in relatively lower concentrations. For balanced optimization (with reasonable yield with improved quality), G500 could be a suitable choice. For applications like water treatment, adsorption or soil amendment, charcoal produced at lower temperatures is recommended, while higher-temperature charcoals are better suited for fuel purposes.

Table 3 compares the fuel properties of charcoal samples with the ISO/TS 17225-:2016(E) standards for thermally treated pellets[54] and the KOR standards for wood charcoal.[55] Compared with the KOR standards, G550 and G600 met the standard for first grade, while G400, G450, and G500 conformed to the second grade for wood charcoal. Meanwhile all samples met the TW2 category, as they only failed to meet the ash content requirement for TW1. However, it is important to note that the parameters specified in ISO/TS 17225-8:2016(E) are specifically intended for graded pellets produced by thermal processing of woody biomass, not for charcoal. ISO/TS 172258:2016(E) provides comprehensive international benchmarks for energy content and compositions at temperatures matching the charcoal grades studied, with pellet-specific criteria (e.g. bulk density, durability) excluded and applied provisionally to charcoal. These facts confirm that Ginkgo charcoal has potential as a solid fuel option, capable of partially substituting for coal in certain applications. Ginkgo charcoal delivers a superior calorific value compared with typical commercial wood charcoals (28–33 MJ kg$^{-1}$) and matches high-rank coals (32–35 MJ kg$^{-1}$),[68,69] with its elevated carbon content, low volatile matter and high fixed carbon more closely resembling anthracitic coals (C 60–90%, VM 10–30%, and FC 60–80%).[68] These attributes underscore its energy-dense profile as a premium solid fuel for cleaner, more efficient combustion, offering superior storage stability and lower emissions compared with conventional wood and lower-rank coals.

Beyond direct energy recovery, converting pruning waste into charcoal extends its life-cycle benefits by displacing fossil fuels and cutting the high emissions from incineration, although further study is needed. Carbon sequestration analysis[70] shows that pyrolyzing Ginkgo retains 46.78% of its original biomass carbon in a stable charcoal form. Compared with conventional disposal, charcoal production prevents roughly 1.4 kg $CO_2$ per kg of woody biomass vs. incineration and about 0.73 kg $CO_2$-equivalent per kg vs. landfilling.[71] Thus, this process delivers a comprehensive waste-to-energy solution that tackles urban waste management while advancing climate mitigation goals.

| Table 3. Fuel properties comparison to charcoal standard. | | | | | | |
|---|---|---|---|---|---|---|
| Standard | Parameter limit | G400 | G450 | G500 | G550 | G600 |
| KOR first-grade charcoal | HHV ≥32.67 MJ kg$^{-1}$ | ✗ | ✗ | ✗ | ✓ | ✓ |
| KOR first-grade charcoal | Ash ≤5% | ✓ | ✓ | ✓ | ✓ | ✓ |
| KOR first-grade charcoal | MC ≤10% | ✓ | ✓ | ✓ | ✓ | ✓ |
| KOR first-grade charcoal | Heavy metals | ✓ | ✓ | ✓ | ✓ | ✓ |
| KOR second-grade charcoal | HHV ≥29.28 MJ kg$^{-1}$ | ✗ | ✗ | ✓ | ✓ | ✓ |
| ISO/TS 17225-8: 2016(E):TW1 | HHV ≥21.00 MJ kg$^{-1}$ | ✓ | ✓ | ✓ | ✓ | ✓ |
| ISO/TS 17225-8: 2016(E):TW1 | Ash ≤1.2% | ✓ | ✗ | ✗ | ✗ | ✗ |
| ISO/TS 17225-8: 2016(E):TW1 | MC ≤8% | ✓ | ✓ | ✓ | ✓ | ✓ |
| ISO/TS 17225-8: 2016(E):TW1 | $N_2$, % ≤0.4% | ✓ | ✓ | ✓ | ✓ | ✓ |
| ISO/TS 17225-8: 2016(E):TW2 | Ash ≤3% | ✓ | ✓ | ✓ | ✓ | ✓ |

*Note*: ✓: Meets the standard; ✗: Does not meet the standard. Other parameters are not applicable or met by default, or have already been discussed in previous sections.
HHV, higher heating value; KOR, Korean regulatory standard; MC, moisture content.

## Conclusion and recommendations

This study focused on production of charcoals from GBP and optimization strategies of pyrolysis for enhancing fuel properties. Through a series of controlled pyrolysis experiments conducted with variation in temperature from 400 to 600°C for 1 h, and then subsequent testing of fuel properties, key findings were obtained regarding the pyrolysis process and its impact on the charcoal characteristics:

1. *Mass and dimension reductions* – the slow pyrolysis process effectively reduced mass by 67.95–73.59%. Ginkgo showed an anisotropic shrinkage pattern for axial and transverse direction with a 41.19–49.97% reduction in approximate volume, indicating energy densified efficient conversion of urban pruning residues into charcoal, facilitating convenient transportation and space saving in storage.
2. The *moisture absorption test* showed a linear negative correlation with the higher processed temperature showing more hydrophobicity, improving features for storage stability and fuel applications.
3. *Fuel property improvement* – Ginkgo charcoal's carbon content rose from 48.33% in the raw material to 85.61% in G600, and its FC from 21.00% to 79.96%. The energy density was improved from 20.76 MJ kg$^{-1}$ (for raw) to 34.26 MJ kg$^{-1}$ (for G600). With the highest carbon content, lowest volatile matter, superior thermal stability, highest FR (6.36) and VI (39.61), and the lowest CI (5.8), the charcoal sample G600 showed excellent fuel properties while having the lowest MY and EY (MY, 26.41%; EY, 43.58%). G400 had the highest MY (32.05%) and EY (46.75%).
4. The charcoal produced demonstrated *upgraded calorific value, favorable elemental composition and compliance with international fuel standards* (ISO and KOR), showing that G550 and G600 met the standard for first grade, while G400, G450 and G500 conformed to the second grade for wood charcoal of Korean standards.
5. *Thermal stability* – charcoals produced at higher temperatures, particularly G600 and G550, exhibited superior thermal stability and combustion performance, with G550 demonstrating the steepest weight loss and best combustion in TGA analysis, making it ideal for high thermal resilience applications.
6. *Heavy metal and trace element concentration* – despite their superior fuel and thermal properties, some charcoals contain higher concentrations of heavy metals, particularly zinc, probably owing to environmental factors, necessitating pretreatment methods or specific applications to ensure safety and mitigate health and environmental risks.

This study demonstrated the potential of GBP as a feedstock for solid fuel (charcoal) production by optimizing pyrolysis temperatures. It highlighted the dual benefits of waste management and renewable energy generation. The findings offer valuable insights into city planning, waste management and policymaking, while also advancing the optimization of pyrolysis processes. Future research should focus on co-pyrolysis with non-woody wastes (e.g. plastics, paper, rubber) to ensure sustainable feedstock availability for urban charcoal production, as well as on reducing heavy metals in charcoal for better environmental protection. Further studies on scalability, bulk collection logistics and long-term environmental impacts are also needed.

## Conflict of interest

The authors declare that they have no known competing financial interests or personal relationships that could have appeared to influence the work reported in this paper.

## Funding information

This research was supported by the Ministry of Science and ICT, Korea, under the Innovative Human Resource Development for Local Intellectualization support program (IITP-2023-RS-2023-00260267) supervised by the Institute for Information and Communications Technology Planning & Evaluation. It was also supported by the National Research Foundation of Korea through a grant funded by the Korea government (Ministry of Science and ICT) (no. 2021R1A6A1A0304424211).

## Data availability statement

The data that support the findings of this study are available from the corresponding author, Professor Kim, upon reasonable request.